\documentclass[english,prl,twocolumn,showpacs,superscriptaddress,floatfix,longbibliography]{revtex4-1}
\pdfoutput=1
\usepackage[T1]{fontenc}
\usepackage[latin9]{inputenc}
\usepackage{amsmath,amssymb,amstext}
\usepackage{xcolor}
\usepackage{graphicx}
\usepackage{esint}
\usepackage{babel}

\makeatletter
\let\c@author\relax
\makeatother

\usepackage{bm,bbold}
\usepackage{color}
\usepackage{subcaption}
\usepackage{dsfont}
\usepackage[labelfont=bf, justification=justified]{caption}

\definecolor{TTH-color}{HTML}{097000}
\definecolor{FSB-color}{named}{magenta}
\definecolor{MR-color}{RGB}{128,0,128}

\definecolor{TTH-color2}{named}{red}
\definecolor{FSB-color2}{named}{blue}
\definecolor{MR-color2}{RGB}{128,0,128}

\begin{document}

\title{Electron refrigeration in hybrid structures with spin-split superconductors}

\author{M. Rouco}
\email{mrouco001@ikasle.ehu.eus}
\affiliation{Centro de F\'isica de Materiales (CFM-MPC), Centro Mixto CSIC-UPV/EHU, Manuel de Lardizabal 5, E-20018 San Sebastian, Spain}
\author{T T Heikkil\"a}
\email{tero.t.heikkila@jyu.fi}
\affiliation{University of Jyvaskyla, Department of Physics and Nanoscience Center,
P.O. Box 35 (YFL), FI-40014 University of Jyv\"askyl\"a, Finland }
\author{F.S. Bergeret \thanks{sebastian_bergeret@ehu.eus}}
\email{sebastian_bergeret@ehu.eus}
\affiliation{Centro de F\'isica de Materiales (CFM-MPC), Centro Mixto CSIC-UPV/EHU, Manuel de Lardizabal 5, E-20018 San Sebastian, Spain}
\affiliation{Donostia International Physics Center (DIPC),Manuel de Lardizabal 4, E-20018 San Sebastian, Spain}

\begin{abstract} 
Electron tunneling between superconductors and normal metals has been used for an efficient refrigeration of electrons in the latter. Such  cooling is a non-linear effect and usually requires a large voltage. Here we study the electron cooling in heterostructures based on superconductors with a spin-splitting field coupled to normal metals via spin-filtering barriers. The cooling power shows a linear term in the applied voltage. This improves the coefficient of performance of electron refrigeration in the normal metal by shifting its optimum cooling to lower voltage, and also allows for cooling the spin-split superconductor by  reverting the sign of the  voltage. We also show how tunnel coupling spin-split superconductors with regular ones allows for a highly efficient refrigeration of the latter.
\end{abstract}

\maketitle

\section{Introduction}

The common way to refrigerate electron systems at sub-Kelvin temperatures is to lower the temperature of the whole sample via different refrigeration methods. In those cases the lattice temperature is lowered first, and the electron-phonon coupling then refrigerates the electrons. This mechanism becomes inefficient at low temperatures as, there, the phonons decouple from the electrons. An alternative scheme is to directly refrigerate the electrons. A scheme for such direct electron refrigeration was presented more than two decades ago \cite{nahum1996,leivo1996}. It is based on electron tunneling between a superconducting (S) electrode and a normal-metal (N) island, where the gapped density of states in S allows for a selective transport of hot electrons out of N by a proper choice of the bias voltage \cite{giazotto2006rmp}. This refrigeration method is very efficient, as the absolute temperature of the N electrons can be lowered to a tiny fraction of the starting temperature \cite{pekola2004,clark2005,rajauria2007,nguyen2014,muhonen2012}. This heat transfer through the junction could be used for the realization of on-chip cooling \cite{nguyen2015} of nanosized systems, such as highly-sensitive detectors and quantum devices. 

At high starting temperatures (in case of Al-based microcoolers, typically above 200 mK) the mechanism limiting the lowest reached electron temperatures is the electron-phonon coupling. However, the electron refrigerators become especially useful below these temperatures. In those cases the limiting factor is rather the low coefficient of power (COP, refrigeration efficiency): as the refrigeration requires electric power which comes with Joule heating, the excess heat is dumped into the superconductor, which then heats up, and the resulting backflow of heat limits the refrigeration. This has been partially cured with a design involving quasiparticle traps \cite{nguyen2014}, but increasing the COP would allow for further progress.

The improvement of the cooling power due to a spin filter between N and superconducting (S) electrodes was theoretically discussed in Ref. \cite{kawabata2013efficient}.  However, the effect of spin-splitting was not considered in that work.  More recently, the electronic cooling power between a ferromagnetic metal and a superconductor in the presence of  an external magnetic field  has been  studied both theoretically and experimentally \cite{Ozaeta2014a,kolenda2016b}.  It was shown that due to the spin-splitting field the cooling power can be larger with respect to the N-S coolers at certain subgap  voltages. This is a  direct consequence of the linear thermoelectric effect predicted in  Ref. \cite{Ozaeta2014a} and first observed in Ref. \cite{Kolenda2016}. Improvement of the cooling at low voltages suggests an improvement of electron refrigeration.  However, the electronic refrigeration and, in particular, the calculation of the reduction of the electron temperature in such structures  has not been reported before. 

Here we propose a way to improve the refrigeration efficiency  by considering spin-split superconductors (SS), {\it i.e.}  superconductors with  a spin splitting in their density of states, coupled to other electrodes via ferromagnetic insulators (FI). The latter, on the one hand, acts as a spin-filter \cite{Moodera_review} and, on the other hand, induces a  spin-splitting in the superconducting electrodes without the need of applying an external magnetic field, as observed in Al/EuS junctions \cite{meservey1970,meservey1975,xiong2011, hao1991, kolenda2017,strambini2017}. In a SS-FI-N junction the cooling power has a linear term in voltage and the optimal cooling power of N shifts to lower voltages. We show that this linear behavior also allows for cooling of the SS as the sign of the bias voltage across the junction is changed. Moreover, we also found that the cooling of the N-electrode in a N-FI-SS junction can be improved if N is substituted by a superconductor S' with a gap smaller than SS gap.  We finally analyze the electron refrigeration by computing  the electron temperature in SS-FI-N-FI-SS, N-FI-SS-FI-N and SS-FI-S'-FI-SS junctions for different voltages and temperatures.

\section{Model and basic  equations}

\begin{figure}[!t]
\includegraphics[width=\linewidth]{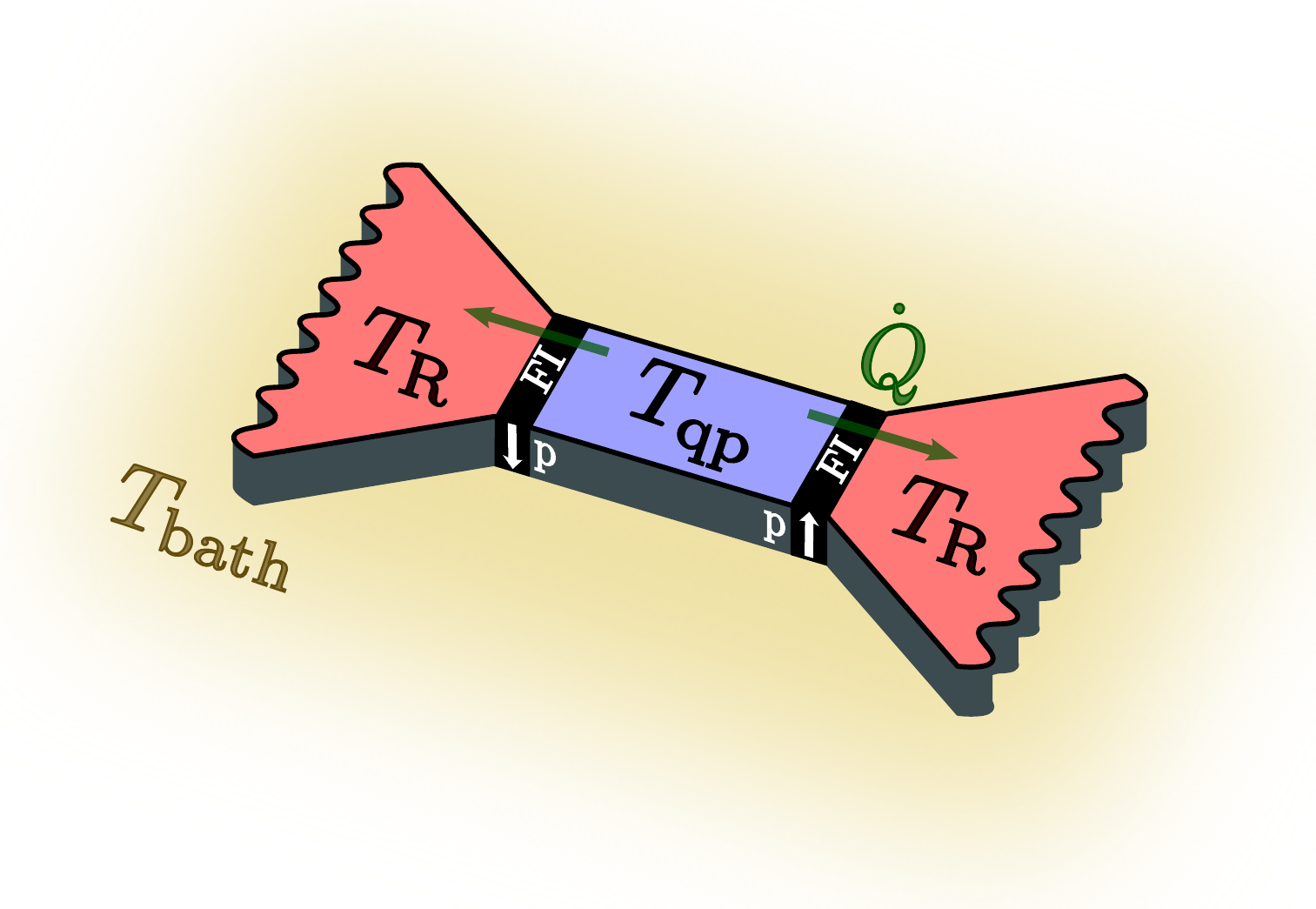}
\caption{Schematic general setup:  An island (blue) with quasiparticle temperature $T_\text{qp}$ is electrically connected to two electrodes with quasiparticle temperature $T_\text{R}$ through a spin-filtering thin layer of a ferromagnetic insulator. We assume  that the film phonons strongly thermalize to the substrate and, thus, their temperature equals $T_\text{bath}$. $\dot{Q}$ stands for the cooling power of the island and white arrows show the spin polarization at each FI.}
\label{fig:system-scheme}
\end{figure}

A typical setup for electron refrigeration is schematically shown in Fig.~\ref{fig:system-scheme}. The central island is the one refrigerated and it is tunnel coupled to two electrodes. The latter can be formed by a normal metal or a spin-split superconductor (SS). Besides the energy exchange between the electrodes and the island, carried by the quasiparticles, electron-phonon coupling establishes an energy flow from quasiparticles to the  phonon bath of the different parts of the system. We assume the film phonons to thermalize strongly to those at the thermal bath (\textit{i.e.}, the substrate), so that their temperature equals $T_\text{bath}$. Decoupling the island from the substrate, however, can lead to phonon refrigeration and a consequent enhancement in the refrigeration of the quasiparticles, as it was reported in \cite{clark2005,nguyen2015}.

We define the cooling power, $\dot{Q}$, as the energy current flowing out of the central island to the electrodes, whereas the energy current flow resulting from  the electron-phonon coupling  is labeled as $\dot{Q}_\text{qp-ph}$. Our goal is to determine the final temperature of the quasiparticles in the central island,  $T_{qp}$, when the system is voltage biased. This temperature can be calculated from the heat balance equation, which in the stationary situation (no build-up of energy) reads
\begin{equation}
\label{eq:quasiequilibrium}
\dot{Q}(T_{qp},T_R) + \dot{Q}_\text{qp-ph}(T_{qp},T_{\rm bath}) = 0\; .
\end{equation}
Here, $T_{R}$ is the quasiparticle temperature in the electrode(s), which can differ from $T_\text{bath}$ if the thermalization is incomplete. In practice, the electrodes may heat up close to the junction. We model this heating by considering a finite size of the electrodes as well as that of the island. In addition, the  cooling power $\dot Q$ depends on  the bias voltage and other system parameters, as discussed in each of the examples consider below.    

The quasiparticle-phonon heat flow entering Eq.~\eqref{eq:quasiequilibrium} is given by  \cite{maisi2013,timofeev09}
\begin{align}
\label{eq:qdot-ep-SS}
  \dot{Q}_\text{qp-ph} = &-\frac{\Sigma \Omega}{96 \zeta(5)}
           \int_{-\infty}^{\infty} dE \; E \int_{-\infty}^{\infty} d\omega \; \omega^2 \;
           {\rm sgn}(\omega) L_{E, E+\omega} \nonumber \\ 
           & \times \bigg\{ \coth\left(\frac{\omega}{2T_{\rm ph}} \right)\left[\tanh\left(\frac{E}{T_{\rm qp}}\right) 
           \tanh\left(\frac{E+\omega}{2T_{\rm qp}}\right)\right] \nonumber \\
           &- \tanh\left(\frac{E}{2T_{\rm qp}}\right) \tanh\left(\frac{E+\omega}{2T_{\rm qp}}\right) + 1 \bigg\}\; , 
\end{align} 
where $\Sigma$ is a constant describing the coupling strength, $\Omega$ stands for the sample volume, $\zeta(5) \approx 1.037$, we set $k_B=1$, and the kernel $L_{E, E'}$  reads
\begin{equation}
  L_{E,E'} = \frac{1}{2} \sum_{\sigma = \pm} N_S(E_\sigma) N_S(E'_\sigma) \left[1 - \frac{\Delta(T_{qp},h)}{E_\sigma E'_\sigma} \right]\;.
\end{equation}
Here $N_{S} (E) = \text{Re}\left[(E+i\Gamma)/\sqrt{(E+i\Gamma)^2-\Delta^{2}}\right]$ is the density of states of the superconductor, $E_\pm \equiv E\pm h$, $h$ is the effective spin splitting and $\Delta(T_{qp},h)$ is the self-consistently \cite{bergeret2017nonequilibrium} calculated superconducting order parameter. If the central conductor is a normal metal (i.e. $\Delta_0 \equiv \Delta(0, 0) = 0$), $\dot{Q}_\text{qp-ph}$ simplifies to \cite{wellstood1994}
\begin{equation}
\label{eq:qdot-ep-N}
\dot{Q}_\text{qp-ph}^N = \Sigma\Omega \left(T_\text{qp}^5 - T_\text{ph}^5 \right).
\end{equation}
These results, along with the ones shown in the following section, are used  in Sec. \ref{sec:results} to determine the final electronic temperature of the island for different systems with geometries equivalent to the one displayed in Fig.~\ref{fig:system-scheme}.

\section{Cooling power of a SS-FI-N junction}
\label{sec:qdot}

We first analyze the cooling power $\dot{Q}$ of a SS-FI-N junction, where SS is a spin-split superconductor.   We assume that the spin-filter efficiency of the ferromagnetic insulator is 100\% and hence we can neglect the Andreev-reflection  processes at the barrier \cite{Ozaeta2012a}. Such an assumption is justified in the case of EuO and EuS barriers with spin-filter efficiencies $>95\%$ \cite{santos2004}.  The suppression of coherent processes such as Andreev reflection enables us to describe the problem within the  tunneling formulation   (see below Eq.~\eqref{eq:qdot})  which substantially simplifies the calculations. 

\begin{figure}[!t]
    \centering
    \includegraphics[width=\linewidth]{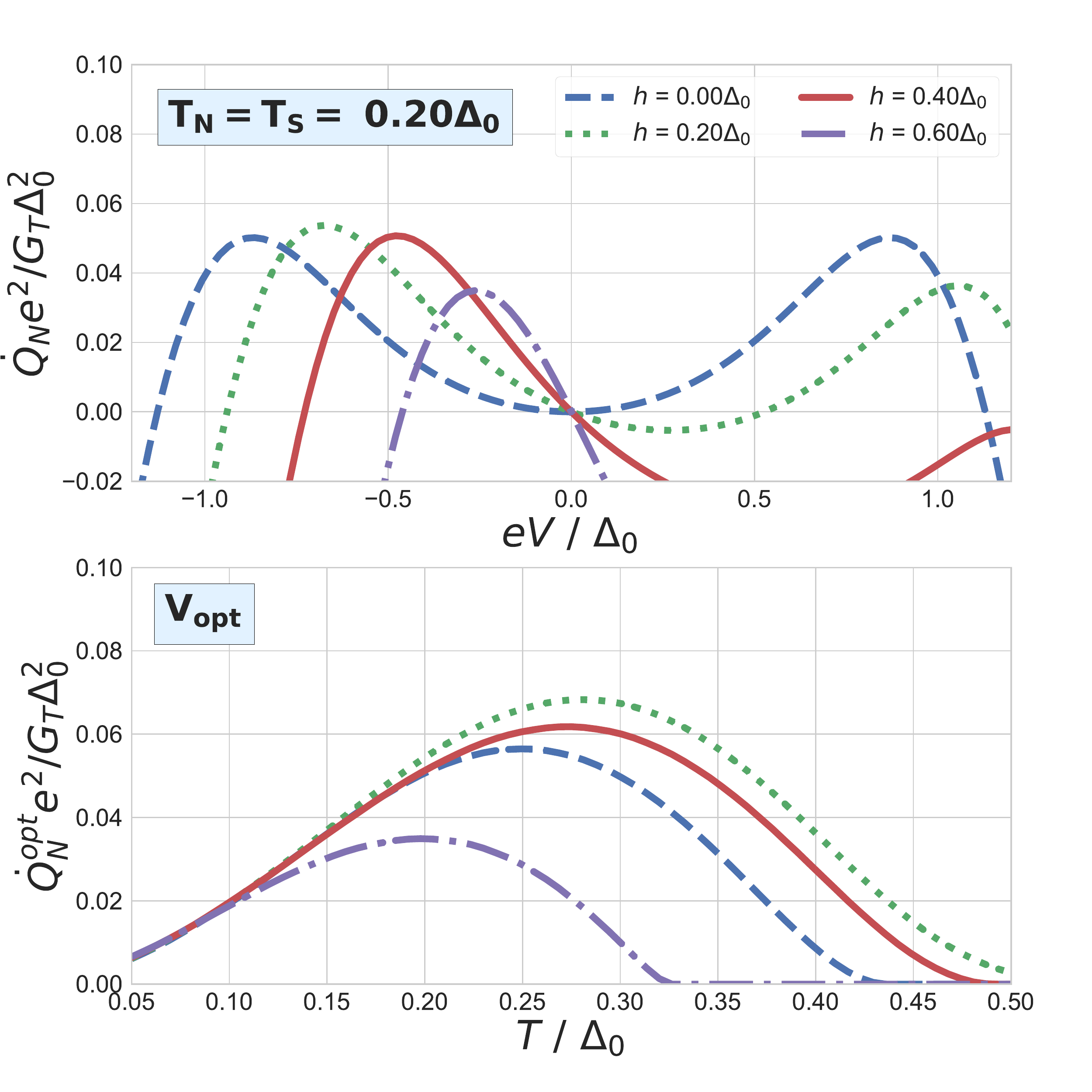}
    \caption{(\textit{top}) Cooling power $\dot Q_N$ of the metallic electrode as a function of the bias voltage at $T_N = T_S = 0.2\Delta_0$ for different values of the exchange field $h$. (\textit{bottom}) Cooling power optimized vs. bias voltage as a function of the temperature of the junction $T = T_N = T_S$. $V_\text{opt}$ stands for the applied bias for which $\dot Q^{opt}$ is obtained.}
    \label{fig:qdot-n}
\end{figure}
\begin{figure}[!t]
    \centering
    \includegraphics[width=\linewidth]{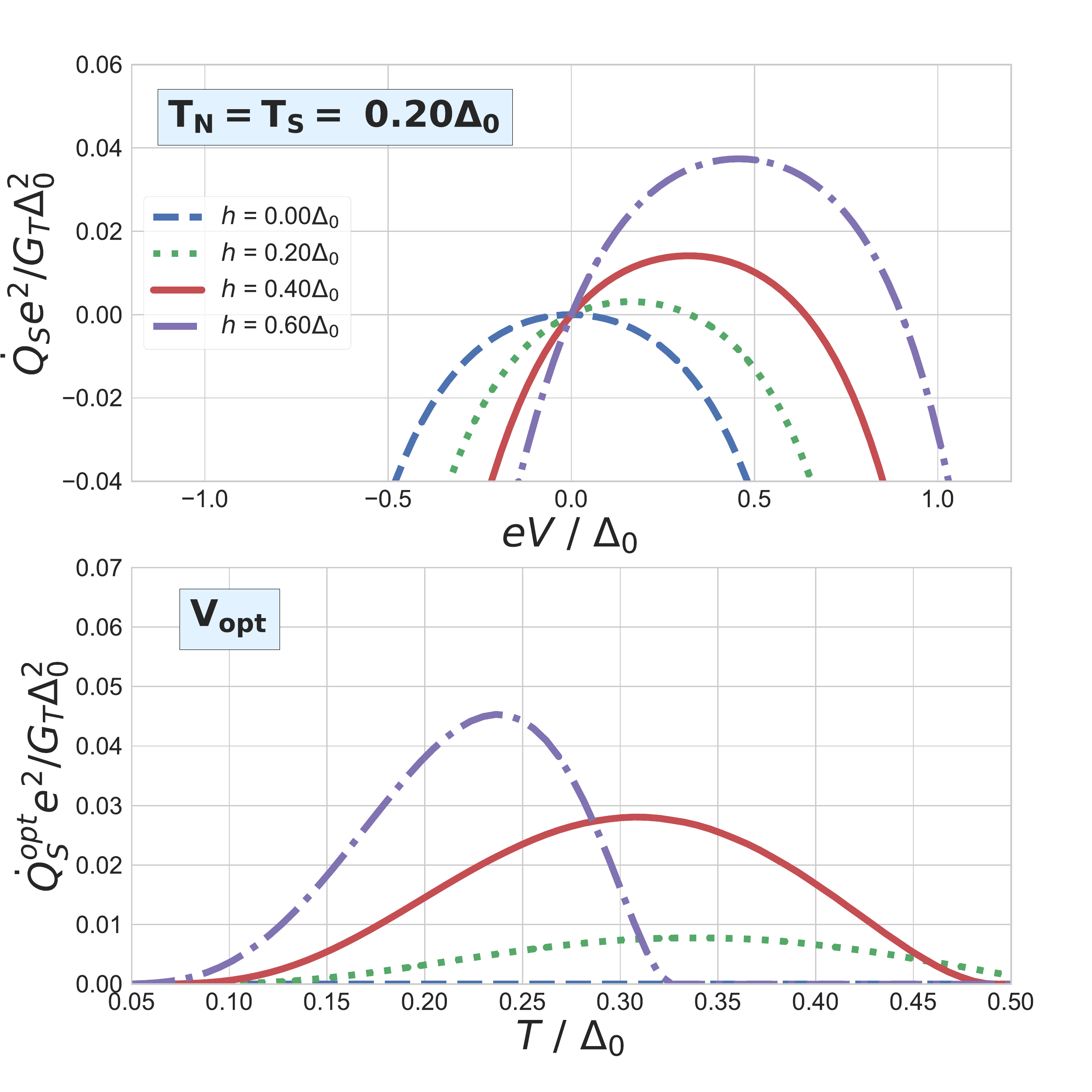}
    \caption{(\textit{top}) Cooling power $\dot Q_S$ of the spin-split superconductor as a function of the bias voltage for different values of the exchange field $h$, at $T_S = T_N = 0.2\Delta_0$. (\textit{bottom}) Cooling power for optimal bias voltage $V_{\rm opt}$, as a function of the temperature of the junction $T = T_N = T_S$.}
    \label{fig:qdot-s}
\end{figure}

The junction can be both temperature and voltage biased, and the cooling power in each of the electrodes is given by \cite{bergeret2017nonequilibrium},
\begin{equation}
\label{eq:qdot}
	\dot{Q}_i = \frac{G_T}{e}\int^\infty_{-\infty} dE(E+\mu_i) N_N \left[N_+ + pN_- \right] (f_{N_i} - f_{SS}),
\end{equation}
where $i=\{N,SS\}$ labels the metal and the spin-split superconductor, $\mu_i = eV_i$ is the electrochemical potential on each of the samples, $G_T$ is the normal-state conductance of the junction, $N_\pm \equiv \tfrac{1}{2}\left[N_S(E+h) \pm N_{S}(E-h)\right]$ takes into account the spin-split DOS of the superconductor and $f_{SS} \equiv f(E)$ and $f_{N_i} \equiv f(E+eV_i)$, where $f(E) = 1 / (1 +e^{E/T})$ stands for the Fermi-Dirac distribution function.

The combination of spin splitting and spin filtering effectively moves the Fermi level in the superconductor from the middle of the superconducting gap an energy equal to $h$. This situation breaks the preexistent electron-hole symmetry of the currents and, consequently, it makes $\dot Q$ asymmetric in voltage.

In Figs.~\ref{fig:qdot-n} and \ref{fig:qdot-s} the cooling power in the normal metal, $\dot{Q}_N$, and the spin-split superconductor, $\dot{Q}_{SS}$,  are shown for different spin-splitting amplitudes $h$. In the upper panels their bias dependence at a given temperature is shown, whereas in the lower panels the maximum of cooling power $\dot Q^{opt}$ is represented in terms of the temperature of the junction.

For the  normal-metal cooling (Fig. \ref{fig:qdot-n}), the most important effect is the shift of the maximum cooling power, $\dot Q^{opt}_N$, towards lower values of $V$ by increasing the value of $h$.   Since  the joule heating driven by dissipative currents is given by $P = IV$, this shift towards lower bias voltages implies lower dissipation \cite{Kolenda2016}. On the other hand the maximum  cooling power  $\dot{Q}_N$ may decrease on increasing $h$ due to the suppression of the superconducting gap.  This is the case of the $h=0.6\Delta_0$ curve in Fig.~\ref{fig:qdot-n}, for which the superconductor makes a transition to the normal state at $T \approx 0.32\Delta_0$. In the next section we study the consequences  of this behavior on  the electronic refrigeration.

If we now focus on the cooling power in  the superconducting electrode (Fig. \ref{fig:qdot-s}) it is interesting to notice that the  spin splitting opens the possibility to refrigerate it. In the upper panel of Fig. \ref{fig:qdot-s} one clearly sees the linear behavior of the cooling power for low positive voltages  for a non-zero exchange field $h$.  In this case, larger  spin-splitting implies better cooling. Of course the superconducting phase transition limits this enhancement only to temperatures lower to the critical one, as can be seen in the bottom panel of Fig.~\ref{fig:qdot-s}.

\section{Electron refrigeration}

\label{sec:results}
\subsection{N-FI-SS-FI-N structure}
\label{sec:nssn}

We first consider a spin-split superconducting island between two metallic electrodes. We assume both junctions to be identical and, in order to enhance the cooling power, the polarizations of the spin filters are directed opposite to each other. Such a setting is analogous to the n-p-n or p-n-p thermoelectric setups containing two junctions with opposite thermoelectric coefficients in series. In this configuration, the total heat current flowing out of the island is  two times larger than  the one obtained from  Eq.~\eqref{eq:qdot} for a single interface.

Cooling power only  gives  information about the extraction of \textit{hot carriers} from the system of interest. In order to quantify  the refrigeration (which implies change in the temperature) thermalization processes must be taken into consideration via the electron-phonon coupling. Combining the cooling power, $\dot Q$, with the heat  flowing to the phonons, $\dot{Q}_\text{qp-ph}$ (Eq.~\eqref{eq:qdot-ep-SS}), we can obtain the final temperatures of the island by solving Eq.~\eqref{eq:quasiequilibrium}. 

In order to simplify the notation, we group all the parameters in a single dimensionless one,
\begin{equation}
\tilde{\Sigma} \equiv \frac{\Sigma \Omega \Delta_0^3 e^2}{G_T k_B^5}.
\end{equation}
Values of the coupling parameter $\Sigma$ are presented in Ref. \cite{giazotto2006rmp}. We set  $\tilde{\Sigma} = 300$ which corresponds to a junction with a resistance of $R_T \sim$ 1 k$\Omega$ in an aluminum sample with $\Omega \sim 1$ ($\mu$m)$^3$.

\begin{figure}[!t]
  \centering
  \includegraphics[width=\linewidth]{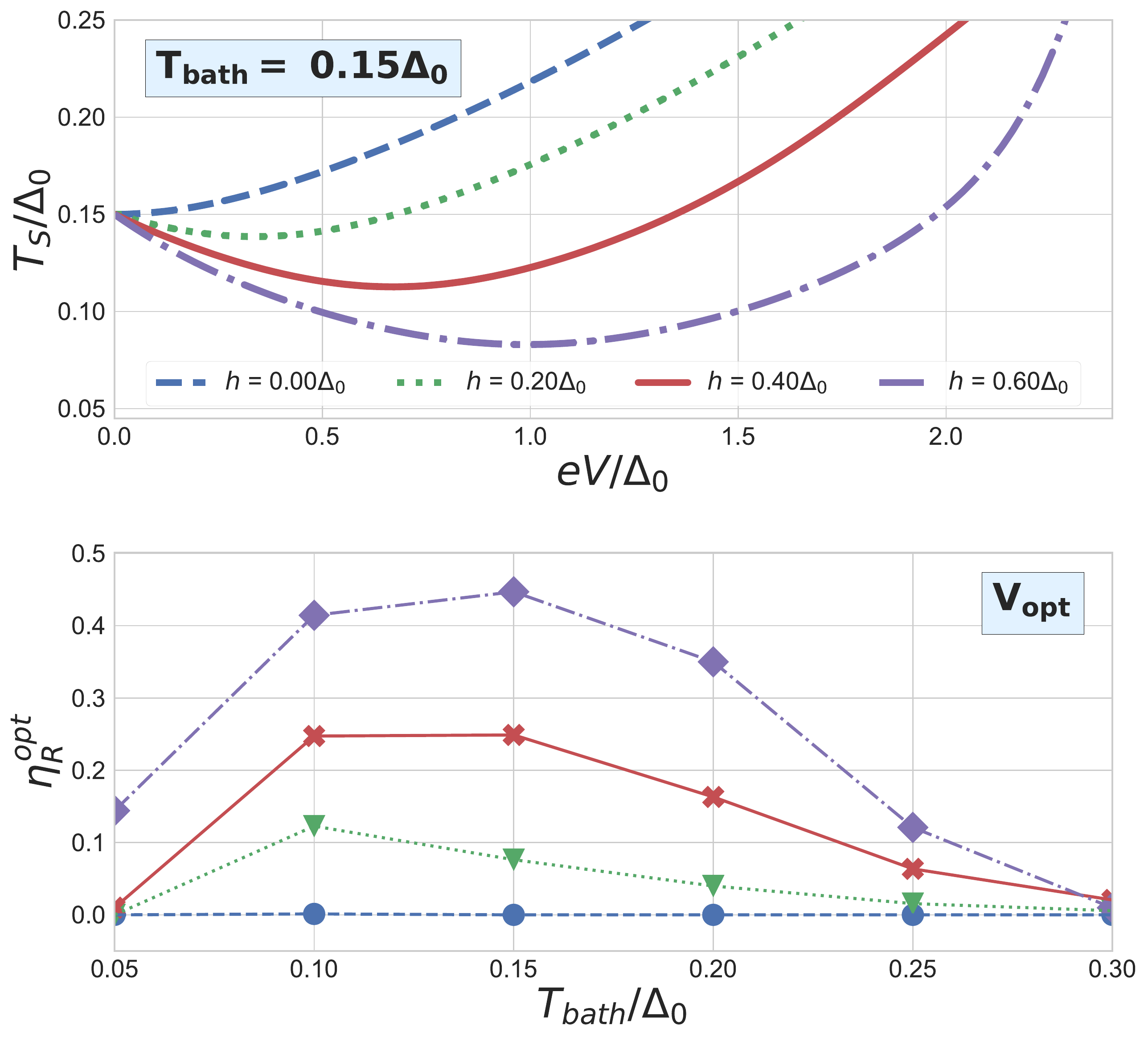}
  \caption{(\textit{top}) Final temperature of the superconducting island vs. the voltage across the structure for $T_\text{bath} = 0.15\Delta_0$. (\textit{bottom}) Optimum value of the relative refrigeration efficiency, $\eta_R \equiv (T_{\rm bath}-T_{qp}) / T_{\rm bath}$, in terms of phonon temperature. Both values are obtained for different spin-splitting amplitudes and a qp-ph interaction parametrized by $\tilde{\Sigma} = 300$.}
  \label{fig:T-super}
\end{figure}

The top panel of Fig.~\ref{fig:T-super} shows the final temperatures of the spin-split superconducting island as a function of the bias voltage across the whole structure for different values of the exchange field $h$. Here, we disregard heating of the electrodes, so the quasiparticle temperature in them equals the bath temperature, $T_N = T_{\rm bath}$. In the absence of an exchange field, refrigeration of the superconducting island cannot be achieved, whereas a non-zero spin-splitting field allows for it.

The bottom panel of Fig.~\ref{fig:T-super} shows  the  refrigeration efficiency at different bath temperatures  $T_{\rm bath}$. For this, we define the \textit{relative refrigeration efficiency},
\begin{equation}
	\label{eq:realtive-ref}
    \eta_R \equiv \frac{T_{\rm bath} - T_{\rm qp}}{T_{\rm bath}},
\end{equation}
indicating the relative temperature drop in the island, and plot it for different $h$ values.
The bottom panel of Fig.~\ref{fig:T-super} shows a strong dependence of the refrigeration efficiency on the temperature of the phonon bath, which is consequence of the competition between how $\dot Q_{SS}^{opt}$ and $\dot{Q}_{\rm qp-ph}^{SS}$ change with $T_{\rm qp}$ and $T_{\rm bath}$. The dependence of the former is shown in Fig.~\ref{fig:qdot-s} and discussed in Sec. \ref{sec:qdot}, while the absolute value of the latter quickly increases with increasing temperature (\textit{i.e.} with an increasing number of phonons). Therefore, at low temperatures the refrigeration efficiency  is mainly governed by $\dot{Q}_{SS}(T_{qp},T_{\rm bath})$, whereas  the rapidly increasing density of phonons governs $\eta^{opt}_R$ at high bath temperatures, dropping the refrigeration efficiency  to very small values.

Notice that at very low temperatures, the electron-phonon relaxation is so weak that, in fact, it does not determine the ultimate minimum temperature that can be achieved. It is the anomalous heating caused by the nonzero DOS within the superconducting gap, which depends on the Dynes parameter $\Gamma$ \cite{dynes1978}, that determines it \cite{pekola2004}.

\begin{figure}[!t]
  \centering
  \includegraphics[width=\linewidth]{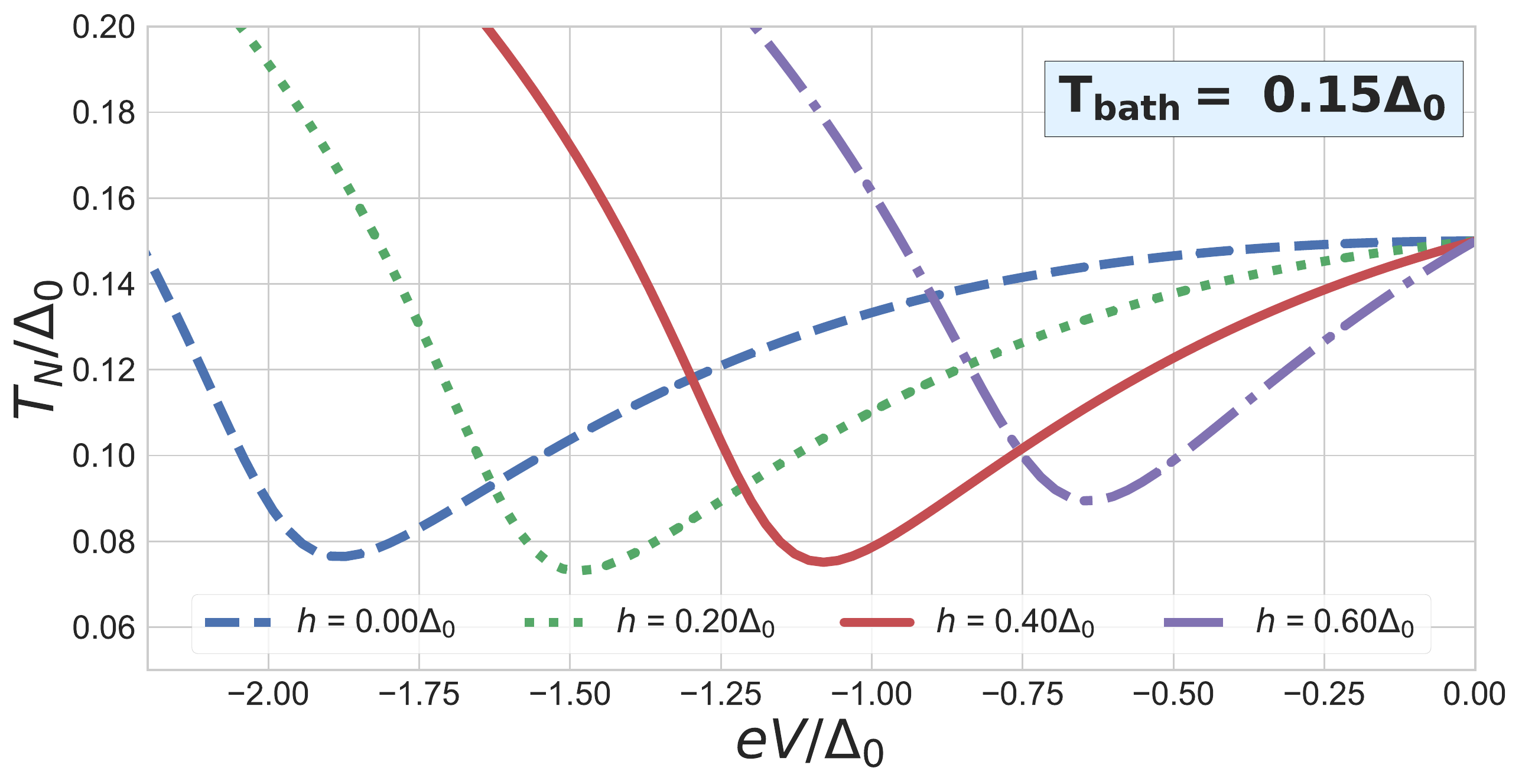}
  \caption{Steady-state temperature in the metallic island attached to two SS electrodes as a function of the bias voltage $V$ across the structure for different exchange fields $h$. Here we disregard the heating of the electrodes. The qp-ph interaction in the normal metal is parametrized by $\tilde{\Sigma}_N=300$.}
  \label{fig:T-normal-inf}
\end{figure}

\subsection{SS-I-N-I-SS structure}
\label{sec:ssnss}

We now focus on the refrigeration of a normal metal island between two spin-split superconducting electrodes. The polarization direction of the spin filters is again opposite to each other in order to optimize the refrigeration. The cooling power in this configuration equals twice the one in Eq.~\eqref{eq:qdot} with $\mu_i = eV/2$, where $V$ is the bias voltage across the whole structure.

As seen in Fig.~\ref{fig:qdot-n}, the spin splitting in the superconducting density of states shifts the maximum of cooling power in the normal metal towards lower values of $V$. This implies lower dissipation and, hence, less heat on the superconducting electrodes.

We first  disregard the heating of the electrodes by assuming  perfect reservoirs for which the temperature  is fixed to $T_\text{bath}$ and  compute the  temperature  $T_N$ in the normal island. Results are shown in Fig.~\ref{fig:T-normal-inf}.

As expected from the results on the cooling power shown in Fig.~\ref{fig:qdot-n}, there is not much  improvement of the refrigeration for a finite $h$, but the optimal refrigeration occurs at lower voltages. 

We now consider the effect of the electrode heating on the previous results.  We do this by adding a second heat balance equation concerning the electrodes, assuming for simplicity that they are uniformly heated across a volume $\Omega_{SS}$. Thus, we get two coupled heat balance equations for the two temperatures $T_N$ and $T_{SS}$, 
\begin{equation}
\label{eq:quasiequilibrium-double}
	\left\{ \begin{array}{l}
		2\dot{Q}_N(T_N, T_{SS}) +  \dot{Q}_\text{qp-ph}^N(T_N) = 0 \\
		\dot{Q}_{SS}(T_N, T_{SS}) +  \dot{Q}_\text{qp-ph}^{SS}(T_{SS}) = 0
	\end{array}\right. ,
\end{equation}
\begin{figure}[!t]
  \centering
  \includegraphics[width=\linewidth]{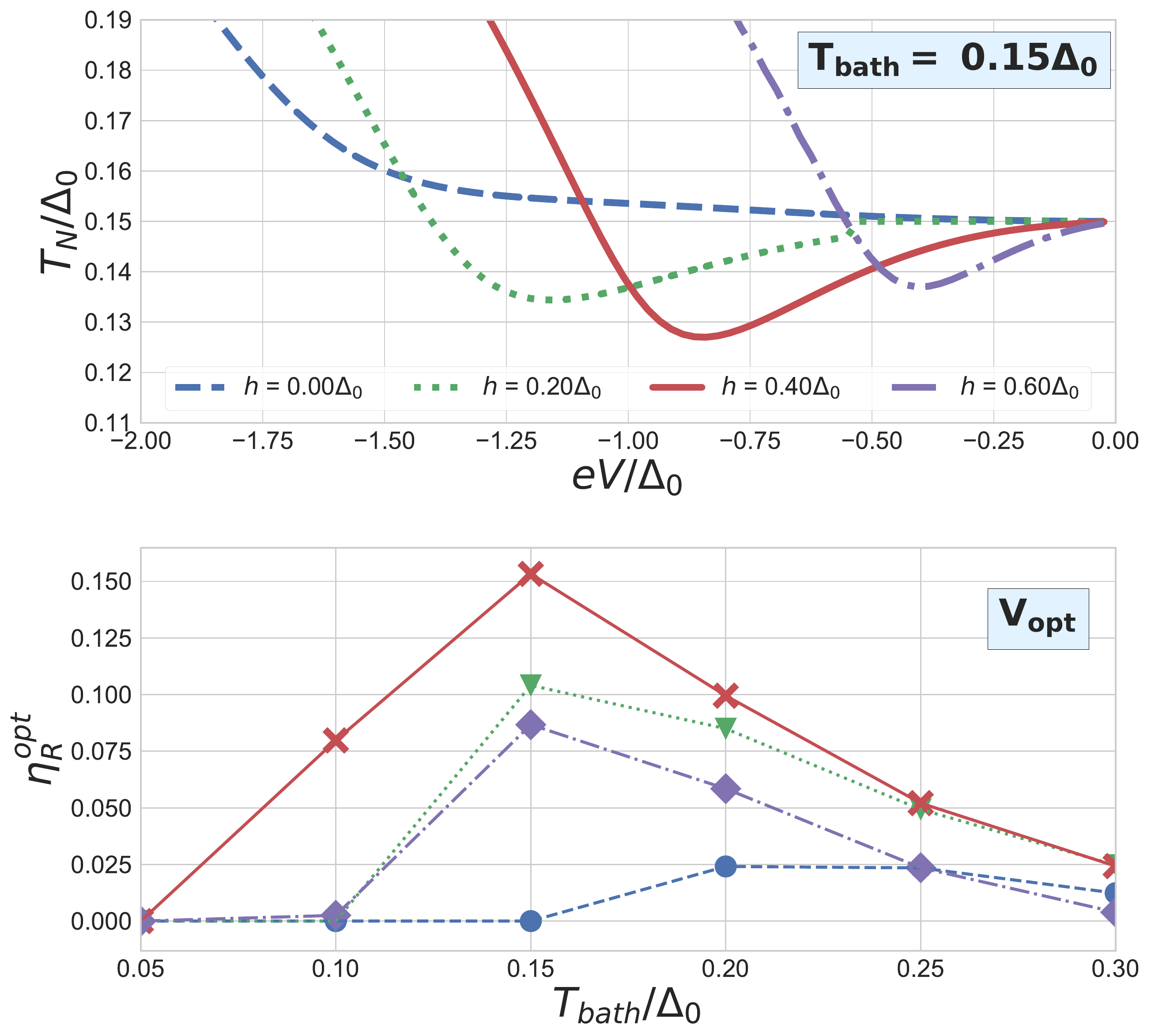}
  \caption{(\textit{top}) Steady-state electronic temperature of the metallic island between two finite spin-split superconductors as a function of the bias voltage $V$ across the whole structure. (\textit{bottom}) Optimum value of the relative refrigeration efficiency as a function of phonon temperature. Results are obtained for different values of $h$ in a particular case where $\tilde{\Sigma}_{SS} = \tilde{\Sigma}_N$.}
  \label{fig:T-normal-finito}
\end{figure}
where the factor of $2$ stems  from the fact that the metallic island is connected to  two electrodes.  Obviously the heat currents in Eq.~\eqref{eq:quasiequilibrium-double} also depend on other parameters like the bias potential or the spin-splitting amplitude, which for simplicity are not explicitly written.

We use Eqs.~\eqref{eq:quasiequilibrium-double} to describe a typical experimental situation. Even though the volume of the electrodes is typically much larger than that of the island, the bad heat transport properties of superconductors at low temperatures would generate a temperature gradient in them, \textit{i.e.} local heating close to the interface.   As shown in Ref. \cite{nguyen2014},  this unwanted effect can be palliated using metallic quasiparticle traps near the junction. The traps dissipate heat more efficiently than the superconductor and, hence, reduces the local temperature of the electrode near the interface. 

In Fig.~\ref{fig:T-normal-finito} we show the obtained results in the case where the volumes of the electrodes and the island are comparable (in particular, when $\tilde{\Sigma}_N = \tilde{\Sigma}_{SS} = 300$). In this configuration, the spin splitting enhances the refrigeration in the island, with an optimal exchange field around $h \sim 0.4\Delta_0$. For higher $h$, the reduction of the Joule heating of the electrodes does not compensate the decrease of $\dot{Q}_S$. Moreover, the crossing between the blue and purple lines in the bottom panel of Fig. \ref{fig:T-normal-finito} is a consequence of the strong decrease of $\Delta$ near the critical temperature. In particular, at $T=0.3\Delta_0$ and $h=0.6\Delta_0$, $\Delta \approx 0.5\Delta_0$.

\subsection{SS-S'-SS structure}
\label{sec:s-ss-s}

\begin{figure}[!t]
	\centering
    \includegraphics[width=\linewidth]{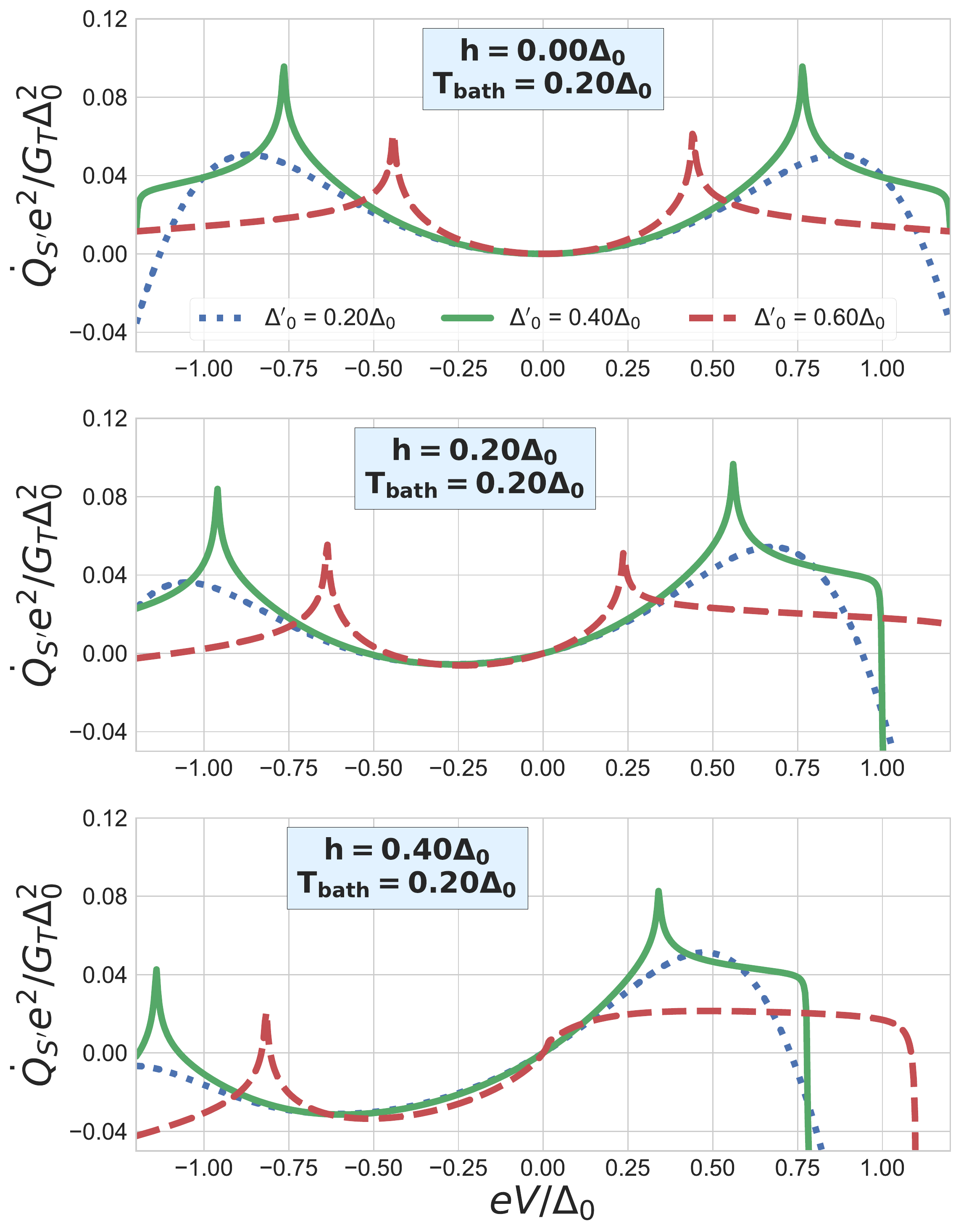}
    \caption{Cooling power $\dot{Q}_{S'}$ as a function of the applied voltage for different exchange fields and order parameters $\Delta_0'=\Delta'(T=0)$ of the superconducting island. $T_{\rm bath} = 0.2\Delta_0$ is set in all the plots.}
	\label{fig:qdot-sprime-nonsplit}
\end{figure}

\begin{figure}[!t]
	\centering
    \includegraphics[width=\linewidth]{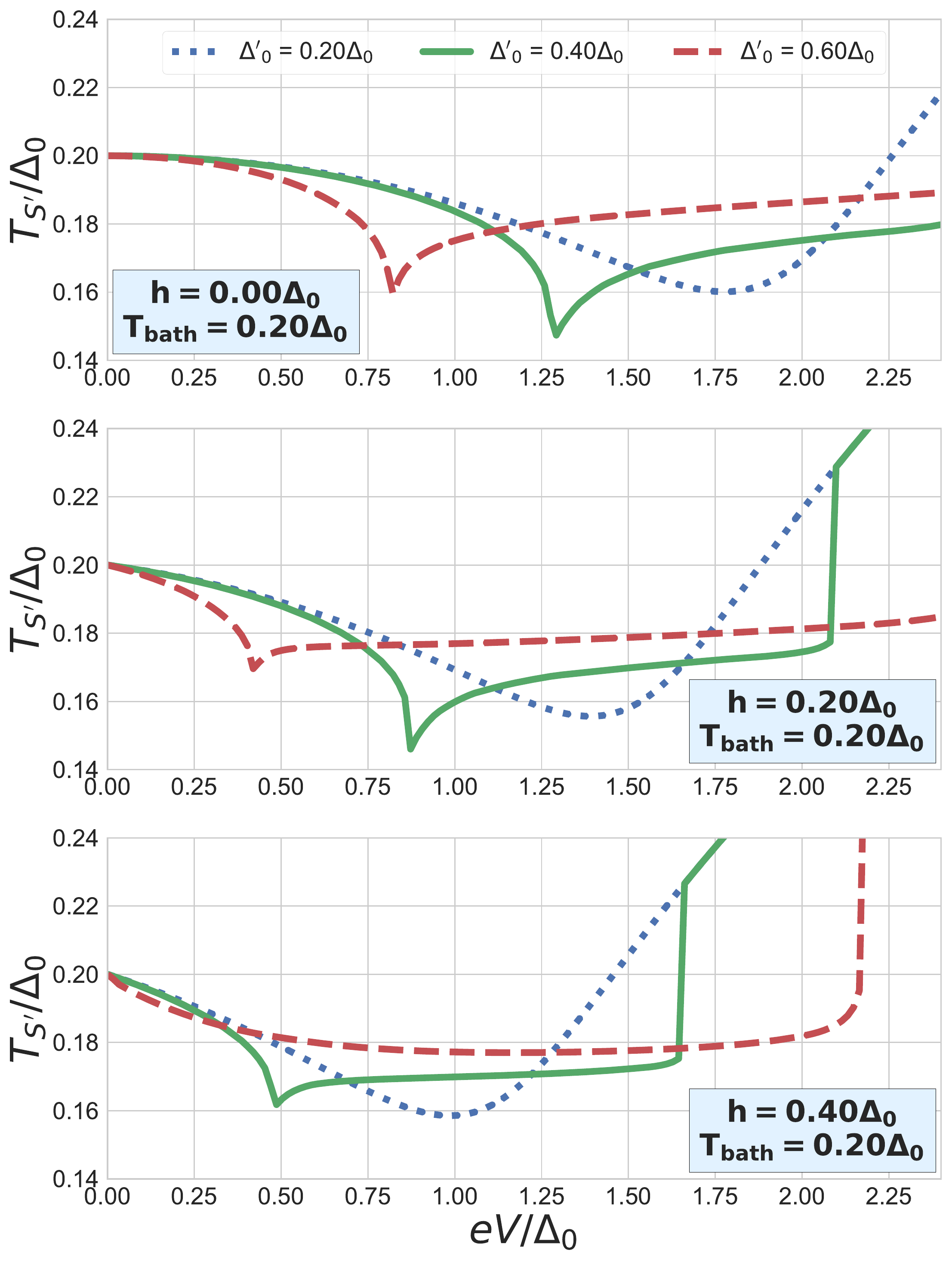}
    \caption{Steady-state temperatures in the island as a function of the bias voltage for different values of exchange fields and superconducting order parameters $\Delta_0'$. We set $T_{\rm bath} = 0.2\Delta_0$, disregard heating of the electrodes, and parametrize the qp-ph interaction in the island by $\Sigma_{S'}=300$.}
    \label{fig:T-sprime-nonsplit}
\end{figure}

We finally  analyze the  refrigeration of  a superconducting island ($S'$) between two spin-split superconducting electrodes ($SS$). The order parameter of the island at zero temperature,  $\Delta_0'$, is assumed to be smaller than the one of the spin-split  electrodes $\Delta_0$ at $T=0$ and $h=0$. 

As in previous cases, we assume two  thin FI layers with opposite polarizations connecting the island with the electrodes.  The cooling power of  the island  is given by Eq.~\eqref{eq:qdot}, where instead of the normal metal DOS we write the usual BCS one: $N_{S'} (E) = \text{Re}\left[(E+i\Gamma)/\sqrt{(E+i\Gamma)^2-\Delta^{'2}}\right]$, where $\Delta' \equiv \Delta'(T_{S'})$ is obtained self-consistently. 

We first show in Fig.~\ref{fig:qdot-sprime-nonsplit} the results  for $\dot{Q}_{S'}$ in a single SS-FI-S' junction, for different values of $h$ and $\Delta_0'$ as a function of the  voltage bias, $V$. The bath temperature has been chosen as $T_{bath}=0.2\Delta_0$. Thus, for  $\Delta_0'=0.2\Delta_0$  (dotted blue curve in Fig.~\ref{fig:qdot-sprime-nonsplit}) the bath temperature is larger than the superconducting critical temperature of the island and, hence, this corresponds to a SS-FI-N junction ({\it cf.}  Fig.~\ref{fig:qdot-n}). For larger values of $\Delta_0'$, the  island is in the superconducting state and its cooling power shows  peaks  at $eV \approx h \pm (\Delta - \Delta')$ (Fig.~\ref{fig:qdot-sprime-nonsplit}), \textit{i.e.} the voltage for which the BCS-coherent peaks in the DOS of S' and SS line up.

Correspondingly, the electron refrigeration is highly enhanced at those voltages, as shown in Fig.~\ref{fig:T-sprime-nonsplit}. In the calculations, we assume that the $SS$ electrodes are perfect  reservoirs with electronic temperature equal to $T_{bath}$. Moreover, in principle a temperature difference between the superconductors may induce a phase-coherent  heat current proportional to $\cos\varphi$, where $\varphi$ is the phase difference between the superconductors \cite{maki1965,giazotto2012}. This current, however, is proportional to $\sqrt{1-p^2}$ and hence vanishes in the case with perfect spin filters \cite{PhysRevB.88.014515} considered here, $p = \pm 1$.

\section{Conclusions}
To conclude, we have shown that superconductors with a spin-split density of states together with spin filters (FI)  may improve the refrigeration  of a metallic (N) island at low voltages.  SS-FI-N junctions also open the possibility to refrigerate the superconductor.
Moreover, if  the N is substituted by a superconductor S' with a gap  smaller than the SS gap,  the refrigeration of S'  can be highly enhanced. These results can be applied to improve current on-chip cooling of metallic components and may lead to many practical applications where the refrigeration of superconductors is 	demanded.

Concerning the material combination for sub-Kelvin coolers,  superconducting aluminum based junctions are the most suitable. Because of the small spin-orbit  interaction, thin Al shows very sharp spin-split density of states  when placed in contact with ferromagnetic insulators, as EuS or EuO \cite{meservey1970, meservey1975, xiong2011, hao1991, kolenda2017,strambini2017}. These europium chalcogenides can also been used as very efficient spin-filtering barriers with  efficiencies of almost 100\% \cite{Moodera_review,Moodera_large_spin_filter}.

\section*{Acknowledgments}
This work was supported by the Spanish Ministerio de Econom\'ia y Competitividad (MINECO) (Project No. FIS2014-55987-P), and the Academy of Finland Center of Excellence (Project No. 284594) and Key Funding (Project No. 305256) programs.

\bibliographystyle{apsrev4-1}
\bibliography{biblist}

\end{document}